\documentclass{article}
\usepackage{amsfonts}
\usepackage{amsmath}

\input{tcilatex}

\begin{document}

\title{$\eta $-weak-pseudo-Hermiticity generators and radially symmetric
Hamiltonians}
\author{Omar Mustafa$^{1}$ and S.Habib Mazharimousavi$^{2}$ \\
Department of Physics, Eastern Mediterranean University, \\
G Magusa, North Cyprus, Mersin 10,Turkey\\
$^{1}$E-mail: omar.mustafa@emu.edu.tr\\
$^{2}$E-mail: habib.mazhari@emu.edu.tr}
\maketitle

\begin{abstract}
A class $\eta $-\emph{weak-pseudo-Hermiticity generators} for spherically
symmetric non-Hermitian Hamiltonians are presented. An operators-based
procedure is introduced so that the results for the 1D Schr\"{o}dinger
Hamiltonian may very well be reproduced. A generalization beyond the
nodeless states is proposed. Our illustrative examples include $\eta $-\emph{%
weak-pseudo-Hermiticity generators} for the non-Hermitian weakly perturbed
1D and radial oscillators, and the non-Hermitian perturbed radial Coulomb.

\end{abstract}

\section{Introduction}

The consensus that \emph{"the existence of real spectrum need not
necessarily be attributed to the Hermiticity of the Hamiltonian"} has
offered a sufficiently strong motivation for the continued interest in the
complex, non-Hermitian, Hamiltonians [1-15]. Intensive studies on such
Hamiltonians resulted in the proposal of the $\mathcal{PT}$ -symmetric
quantum mechanics by Bender and Boettcher [2], where the Hamiltonian
Hermiticity assumption $H=H^{\dagger }$ is replaced by the mere $\mathcal{PT}
$ -symmetry, where $\mathcal{P}$ denotes the parity ($\mathcal{P}x\mathcal{P}%
=-x$) and the anti-linear operator $\mathcal{T}$ mimics the time reflection (%
$\mathcal{T}i\mathcal{T}=-i$). Among the first $\mathcal{PT}$ -symmetric
models with physically acceptable impact has been the Buslaev and Grecchi
quartic anharmonic oscillator [1] described by the radial Schr\"{o}dinger
equation%
\begin{equation}
\left\{ -\frac{d^{2}}{dr^{2}}+\frac{\ell \left( \ell +1\right) }{r^{2}}%
+V\left( r\right) \right\} \psi \left( r\right) =E\psi \left( r\right) .
\end{equation}%
Where, in the presence of a less traditional context, a coordinate shift to
the complex plane, $r\rightarrow x-ic$, is introduced (with real $x\in
\left( -\infty ,\infty \right) $ and a constant $\func{Im}r=-c<0$) and $\psi
\left( r\right) \in L_{2}\left( -\infty ,\infty \right) $ is required. For
explicit illustration the reader may refer to, e.g., Znojil and L\'{e}vai in
[1].

However, subsequent recent studies emphasized that the $\mathcal{PT}$%
-symmetric Hamiltonians constitute a subclass of a very broader class of the
so-called pseudo-Hermiticity of these non-Hermitian Hamiltonians [4-8]. A
Hamiltonian $H$ is pseudo-Hermitian if \thinspace it obeys the similarity
transformation:%
\begin{equation}
\eta \,H\,\eta ^{-1}=H^{\dagger }
\end{equation}%
Where $\eta $ is a Hermitian (and so is $\eta \,H$) invertible linear
operator and $(^{\dagger })$ denotes the adjoint. In such settings, it is
concreted (cf, e.g. [5,7,8]) that $H$, with a complete biorthonormal
eigenvectors, has a real spectrum and is $\eta $-\emph{pseudo-Hermitian}
with respect to the nontrivial "metric"operator 
\begin{equation}
\eta =O^{\dagger }O.
\end{equation}%
Where $O$ is some linear invertible operator $O:\mathcal{H}{\small %
\rightarrow }\mathcal{H}$, and $\mathcal{H}$ is the Hilbert space of the
quantum system with a Hamiltonian $H$ and infinitely many $\eta $ satisfying
(2).(cf, e.g., [4,8,9]). Moreover, an $\eta $-\emph{pseudo-Hermitian}
Hamiltonian equivalently satisfies the \emph{intertwining} relation:%
\begin{equation}
\eta \,H=H^{\dagger }\,\eta .
\end{equation}

However, one may relax $H$ to be $\eta $-\emph{weak-pseudo-Hermitian} by not
restricting the intertwining \emph{second-order} differential operator $\eta 
$ to be Hermitian (cf., e.g., Bagchi and Quesne [9]), linear and/or
invertible (cf. e.g., Solombrino [9]). Yet, without enforcing invertibility
as a necessary condition on $O$ and hence on $\eta $, Fityo [9] has
implicitly used $\eta $-\emph{weak-pseudo-Hermiticity} and constructed 1D
non-Hermitian $\eta $-\emph{weak-pseudo-Hermitian Hamiltonians} via 1D $\eta 
$-\emph{weak-pseudo-Hermiticity generators.} Moreover, very recently we have
(Mustafa and Mazharimousavi [9]) considered the nodeless states of some
non-Hermitian $d$-dimensional Hamiltonians with position-dependent mass and
their $\eta $-(\emph{weak)-pseudo-Hermiticity generators}. We have also
explored [9] exact solvability as a byproduct of such generators in 1D.

In this work, we follow Fityo's [9] ( as well as our work in [9]) $\eta $-%
\emph{weak-pseudo-Hermiticity} condition (i.e., not enforcing invertibility
condition on $O$ and hence on $\eta $) and generalize it to present a class
of spherically symmetric non-Hermitian $\eta $-\emph{weak-pseudo-Hermitian
Hamiltonians} via their $\eta $-\emph{weak-pseudo-Hermiticity generators},
for multi-nodal states. An operators-based procedure is introduced (in
section 2) so that the results of Fityo's formalism [9], for the
one-dimensional (1D) Schr\"{o}dinger Hamiltonian, may very well be
reproduced. On this issue, the reader may be reminded that one can safely
return back from the redial Schr\"{o}dinger equation (1) to the one
dimensional case by the trivial choice $\ell =-1$ and/or $\ell =0$ (cf,
e.g., the sample in [3]). Our illustrative examples, in section 3, include $%
\eta $-\emph{weak-pseudo-Hermiticity generators} for the non-Hermitian
weakly perturbed 1D and radial oscillators, and the non-Hermitian perturbed
radial Coulomb models. Therein, we present not only nodeless radial wave
functions but also some of the multi-nodal ones. We give our concluding
remarks in section 4.

\section{$\protect\eta $-\emph{weak-pseudo-Hermiticity generators} and
radial symmetry; a second-order intertwining $\protect\eta $}

Consider a class of spherically symmetric non-Hermitian Hamiltonians (in $%
\hbar =2m=1$ units) of the form%
\begin{equation}
H=-\partial _{r}^{2}+V_{eff}\left( r\right) ;\text{ \ \ }V_{eff}\left(
r\right) =\frac{\ell \left( \ell +1\right) }{r^{2}}+V\left( r\right)
+iW\left( r\right) ,
\end{equation}%
where \thinspace $V\left( r\right) $ and $W\left( r\right) $ are real-valued
radial functions, and $\ell $ is the angular momentum quantum number. Then $%
H $ has a real spectrum if there is a linear operator $O:$ $\mathcal{H}%
\longrightarrow \mathcal{H}$ such that $H$ is an $\eta $-\emph{%
weak-pseudo-Hermitian Hamiltonian} satisfying the intertwining relation (4).
With the linear operator%
\begin{equation}
O=\partial _{r}+Z(r)\implies O^{\dagger }=-\partial _{r}+Z^{\ast }(r)
\end{equation}%
where 
\begin{equation}
Z(r)=F(r)+iG(r),\text{ \ \ }F(r)=-\left( \ell +1\right) /r+f\left( r\right)
\end{equation}%
and $F(r)$ and $G(r)$ are real-valued radially symmetric functions and $%
\mathbb{R}
\ni r\in \left( 0,\infty \right) $, equation (3) implies%
\begin{equation}
\eta =-\partial _{r}^{2}+M(r)\partial _{r}+N(r),
\end{equation}%
where $M(r)=Z^{\ast }(r)-Z(r)$, $N(r)=Z^{\ast }(r)Z(r)-Z^{\prime }\,(r)$,
and primes denote derivatives with respect to $r.$ Herein, it should be
noted that the operators $O$ and $O^{\dag }$ are two intertwining operators
and the Hermitian operator $\eta $ only plays the role of a certain
auxiliary transformation of the dual Hilbert space and leads to the
intertwining relation (4) (cf, e.g.,[4, 5-10]). Hence, using relation (4)
along with the eigenvalue equation for the Hamiltonian, $H/E_{i}\rangle
=E_{i}/E_{i}\rangle $, and its adjoint, $H^{\dag }/E_{i}\rangle =E_{i}^{\ast
}/E_{i}\rangle $, one can show that any two eigenvectors of $H$ satisfy (cf,
e.g., Mostafazadeh in [4]) 
\begin{equation}
\langle E_{i}/H^{\dagger }\eta -\eta H/E_{j}\rangle =0\implies \left(
E_{i}^{\ast }-E_{j}\right) \,\langle \langle E_{i}/E_{j}\rangle \rangle
_{\eta }=0\,.
\end{equation}%
Which implies that if $E_{i}^{\ast }\neq E_{j}$ then $\langle \langle
E_{i}/E_{j}\rangle \rangle _{\eta }=0.$Therefore, the $\eta $- orthogonality
of the eigenvectors suggests that if $\psi $ is an eigenvector (of
eigenvalue $E=E_{1}+iE_{2},$ $\forall E_{1},E_{2}\in 
\mathbb{R}
$) related to $H$ then%
\begin{equation}
\eta \psi =0\implies O^{\dag }O\psi =0\implies O\psi =0,
\end{equation}%
and%
\begin{equation}
Z(r)=-\frac{\psi ^{\prime }\left( r\right) }{\psi \left( r\right) }%
=-\partial _{r}\ln \psi \left( r\right) \implies \psi \left( r\right) =\exp
\left( -\int^{r}Z(z)dz\right) .
\end{equation}%
Let us recast the linear operators' proposal in (6) as%
\begin{equation}
\partial _{r}=O-Z(r)\text{ \thinspace \thinspace }\implies -\partial
_{r}=O^{\dagger }-Z^{\ast }(r),
\end{equation}%
to imply 
\begin{equation*}
-\partial _{r}^{2}=O^{\dag }O-O^{\dag }Z(r)-Z^{\ast }(r)O+Z^{\ast }(r)Z(r).
\end{equation*}%
Hence, with $E=E_{1}+iE_{2}$ and $H$ in (5) the eigenvalue problem $H\psi
\left( r\right) =E\psi \left( r\right) $ implies 
\begin{equation}
\left[ O^{\dag }O-O^{\dag }Z(r)-Z^{\ast }(r)O+Z^{\ast }(r)Z(r)+V_{eff}\left(
r\right) \right] \psi \left( r\right) =E\psi \left( r\right) .
\end{equation}%
This in turn, collapses into Riccati-type equation:%
\begin{equation}
Z^{\prime }(r)-Z^{2}\left( r\right) +V_{eff}\left( r\right) =E.
\end{equation}%
The real part of which reads%
\begin{equation}
\func{Re}V_{eff}(r)=-F^{\prime }(r)+F\left( r\right) ^{2}-G\left( r\right)
^{2}+E_{1},
\end{equation}%
and the imaginary part reads\vspace{0pt}%
\begin{equation}
\func{Im}V_{eff}(r)=-G^{\prime }(r)+2F\left( r\right) G\left( r\right)
+E_{2},
\end{equation}

On the other hand, the intertwining relation $H^{\dagger }\,\eta =\eta \,H$
along with (6) and (7) would lead to%
\begin{equation}
\left[ V_{eff}^{\ast }(r)-V_{eff}(r)\right] =-2M^{\prime }(r)\implies \func{%
Im}V_{eff}(r)=-iM^{\prime }(r)
\end{equation}%
\begin{equation}
2V_{eff}^{\prime }(r)=M^{\prime \prime }(r)+2N^{\prime }(r)+\left[ M(r)^{2}%
\right] ^{\prime }
\end{equation}%
and%
\begin{equation}
-V_{eff}^{\prime \prime }(r)+M(r)V_{eff}^{\prime }(r)=-N^{\prime \prime
}(r)-2M^{\prime }(r)N(r)
\end{equation}%
to imply, respectively,%
\begin{equation}
\func{Im}V_{eff}(r)=-2G^{\prime }(r),
\end{equation}%
\begin{equation}
\func{Re}V_{eff}(r)=F\left( r\right) ^{2}-G\left( r\right) ^{2}-F^{\prime
}(r)+\beta
\end{equation}%
and%
\begin{equation}
F\left( r\right) ^{2}-F^{\prime }\left( r\right) =\frac{2G\left( r\right)
G^{\prime \prime }\left( r\right) -G^{\prime }\left( r\right) ^{2}+\alpha }{%
4G\left( r\right) ^{2}},
\end{equation}%
where $\alpha ,\beta \in 
\mathbb{R}
$ are integration constants. Eventually, with (21) and (15) implying 
\begin{equation*}
E_{1}=\beta ,
\end{equation*}%
one would recast (15) as%
\begin{equation}
\func{Re}V_{eff}(r)=\left[ \frac{2G\left( r\right) G^{\prime \prime }\left(
r\right) -G^{\prime }\left( r\right) ^{2}+\alpha }{4G\left( r\right) ^{2}}%
\right] -G\left( r\right) ^{2}+\beta .
\end{equation}%
Moreover, equations (20) and (16) would imply%
\begin{equation}
F\left( r\right) =-\frac{\left[ G^{\prime }(r)+E_{2}\right] }{2G(r)}%
\Longrightarrow f\left( r\right) =\frac{\left( \ell +1\right) }{r}-\frac{%
\left[ G^{\prime }(r)+E_{2}\right] }{2G(r)},
\end{equation}%
which when substituted in (22) yields 
\begin{equation*}
E_{2}^{2}=\alpha .
\end{equation*}%
Obviously, one would accept 
\begin{equation*}
\mathbb{R}
\ni \alpha \geq 0\implies 
\mathbb{R}
\ni E_{2}=\pm \sqrt{\alpha }=\omega \sqrt{\alpha },
\end{equation*}%
and negate 
\begin{equation*}
\alpha <0\implies E_{2}\in 
\mathbb{C}%
\end{equation*}%
since $%
\mathbb{R}
\ni E_{2}\notin 
\mathbb{C}
$, as defined early on. Yet $E_{2}\in 
\mathbb{C}
$ contradicts with the real/imaginary descendants, (15) and (16), of the
Riccati-type equation (14). Therefore, with $E_{2}=\omega \sqrt{\alpha },$%
\begin{equation}
f\left( r\right) =\frac{\left( \ell +1\right) }{r}-\frac{\left[ G^{\prime
}(r)+\omega \sqrt{\alpha }\right] }{2G(r)},
\end{equation}%
and%
\begin{equation}
\psi \left( r\right) =\,\sqrt{G(r)}\exp \left( \int^{r}\left[ \frac{\omega 
\sqrt{\alpha }}{2G(z)}-iG(z)\right] dz\right) .
\end{equation}%
where%
\begin{equation}
G\left( r\right) =g(r)r^{2(\ell +1)}\left[ \left(
r^{n_{r}}+\dsum\limits_{p=0}^{n_{r}-1}A_{p,\ell }^{\left( n_{r}\right)
}\,r^{p}\right) ^{2}\right]
\end{equation}%
Nevertheless, with $\alpha =0$ one can express $G(r)$ in terms of $F(r),$
i.e., 
\begin{equation}
G(r)=\exp \left( -2\int^{r}F(z)\,dz\right) =r^{2\left( \ell +1\right) }\exp
\left( -2\int^{r}f\left( z\right) \,dz\right) .
\end{equation}

Hence, $G\left( r\right) $ and/or $F\left( r\right) $ (equivalently%
\thinspace\ $f\left( r\right) $ and/or $g\left( r\right) ,$ i.e. the
generators reported by Fityo in [9]) can be considered as generating
function(s) of the $\eta $\emph{-weak-pseudo-Hermiticity }of non-Hermitian
Hamiltonians with real spectra, where $\psi \left( r\right) $ in (26) is an
eigen function of $H$ in (5), but not necessarily normalizable. Moreover,
the reality of the spectrum, in the forthcoming sections of our proposal, is
secured by the choice $%
\mathbb{R}
\ni \alpha =0$. It should be noted here that the dependence of the radial
wave function on the orbital angular momentum quantum number $\ell $ is
obvious in (26) and (27). Yet, the formalism of the 1D case described by
Fityo [9] may very well be reproduced by the trivial traditional setting $%
\ell =-1,n_{r}=0$, and $r\rightarrow x\in \left( -\infty ,\infty \right) $.

\section{Illustrative examples}

In this section, we construct $\eta $\emph{-weak-pseudo-Herrmiticity} of
some non-Hermitian Hamiltonians using the above mentioned procedure through
the following illustrative examples:

\subsection{Perturbed 1D-harmonic oscillator $\protect\eta $%
-weak-pseudo-Hermiticity generator(s)}

In this subsection we consider the perturbed 1D-harmonic oscillator (i.e., $%
\ell =-1,\,r\rightarrow x\in \left( -\infty ,\infty \right) $) with its
real-valued generating-function $g\left( x\right) =\exp \left( -x^{2}\right) 
$ in (27) to get:

\begin{description}
\item[A)] For $n_{r}=0,$ the effective potential%
\begin{equation}
V_{eff}\left( x\right) =x^{2}-e^{-2x^{2}}+4ix\,e^{-x^{2}}+\beta -1\text{,}
\end{equation}%
which in turn leads to a $\mathcal{PT-}$symmetric $\eta $\emph{%
-weak-pseudo-Hermitian Hamiltonian} of the form 
\begin{equation}
H=-\partial _{x}^{2}+x^{2}+e^{-x^{2}}\left( 4ix-e^{-x^{2}}\right) +\beta -1,
\end{equation}%
with a corresponding node-less (i.e., $n_{r}=0$) wave function%
\begin{equation}
\psi _{0}\left( x\right) =N_{0}\exp \left( -\frac{1}{2}x^{2}-\frac{1}{2}i%
\sqrt{\pi }\func{erf}\left( x\right) \right) .
\end{equation}%
It should be noted that $\psi _{0}\left( x\right) \sim e^{-x^{2}/2}$ in (31)
represents the exact ground state wave function of the well known
1D-harmonic oscillator. The effective potential, on the other hand,
represents a dominating real1D-harmonic oscillator potential $V\left(
x\right) =x^{2}$ perturbed by a weak interaction, with a real part $\left(
-e^{-2x^{2}}\right) $ and an imaginary part $\left( +4xe^{-x^{2}}\right) $.

\item[B)] For $n_{r}=1,$ and $A_{0,\ell }^{\left( 1\right) }=0,$ the
effective potential%
\begin{equation}
V_{eff}\left( x\right) =x^{2}-x^{4}e^{-2x^{2}}+4ix\,e^{-x^{2}}\left(
x^{2}-1\right) +\beta -3\text{,}
\end{equation}%
which in turn leads to an $\eta $\emph{-weak-pseudo-Hermitian Hamiltonian}
of the form 
\begin{equation}
H=-\partial _{x}^{2}+x^{2}-x^{4}e^{-2x^{2}}+4ix\,e^{-x^{2}}\left(
x^{2}-1\right) +\beta -3,
\end{equation}%
with a corresponding one-nodal (i.e., $n_{r}=1$) wave function%
\begin{equation}
\psi _{1}\left( x\right) =N_{1}\,x\exp \left( -\frac{1}{2}x^{2}+\frac{ix}{2}%
e^{-x^{2}}-\frac{1}{4}i\sqrt{\pi }\func{erf}\left( x\right) \right) .
\end{equation}%
It should be noted that $\psi _{1}\left( x\right) \sim xe^{-x^{2}/2}$ is the
exact first-excited state wave function of the 1D-harmonic oscillator.

\item[C)] For $n_{r}=2,$ $A_{0,\ell }^{\left( 2\right) }=0,$ and $A_{1,\ell
}^{\left( 2\right) }=-1$ the effective potential%
\begin{eqnarray}
V_{eff}\left( x\right) &=&x^{2}-\frac{2}{x}-x^{4}\left( x-1\right)
^{4}e^{-2x^{2}}  \notag \\
&&+4ix\,e^{-x^{2}}\left( x-1\right) \left( x^{3}-x^{2}-2x+1\right) +\beta -5%
\text{,}
\end{eqnarray}%
which in turn leads to an $\eta $\emph{-weak-pseudo-Hermitian Hamiltonian}
of the form 
\begin{eqnarray}
H &=&-\partial _{x}^{2}+x^{2}-\frac{2}{x}-x^{4}\left( x-1\right)
^{4}e^{-2x^{2}}  \notag \\
&&+4ix\,e^{-x^{2}}\left( x-1\right) \left( x^{3}-x^{2}-2x+1\right) +\beta -5,
\end{eqnarray}%
with a corresponding two-nodal (i.e., $n_{r}=2$) wave function%
\begin{eqnarray}
\psi _{2}\left( x\right) &=&N_{2}\,x\left( x-1\right) \exp \left( -\frac{1}{2%
}x^{2}\right)  \notag \\
&&\times \exp \left( ie^{-x^{2}}\left[ \frac{x^{3}}{2}-x^{2}+\frac{5x}{4}-1%
\right] -\frac{5}{8}i\sqrt{\pi }\func{erf}\left( x\right) \right) .
\end{eqnarray}%
\newline
It should be noted that $\psi _{1}\left( x\right) \sim x\left( x-1\right)
e^{-x^{2}/2}$ is the exact first-excited state wave function of the harmonic
oscillator.
\end{description}

\subsection{Perturbed radial harmonic oscillator $\protect\eta $\emph{%
-weak-pseudo-Hermiticity generator(s)}}

With $\ell \geq 0$ and $\,r\in \left( 0,\infty \right) $, we consider the
radial harmonic oscillator generating function $g\left( r\right) =\exp
\left( -r^{2}\right) $ to yield:

\begin{description}
\item[A)] For $n_{r}=0,$ the effective potential%
\begin{eqnarray}
\func{Re}V_{eff}\left( r\right) &=&\frac{\ell \left( \ell +1\right) }{r^{2}}%
+r^{2}-r^{4\left( \ell +1\right) }e^{-2r^{2}}-2\left( \ell +1\right) +\beta
-1  \notag \\
\func{Im}V_{eff}\left( r\right) &=&4r^{\left( 2\ell +1\right) }\left[
r^{2}-\left( \ell +1\right) \right] e^{-r^{2}}
\end{eqnarray}%
which in turn leads to an $\eta $\emph{-weak-pseudo-Hermitian Hamiltonian}
of the form 
\begin{equation}
H=-\partial _{r}^{2}+\func{Re}V_{eff}\left( r\right) +i\func{Im}%
V_{eff}\left( r\right) ,
\end{equation}%
with a corresponding node-less (i.e., $n_{r}=0$) wave function%
\begin{equation}
\psi _{0,\ell }\left( r\right) =N_{0,\ell }\,r^{\left( \ell +1\right) }\exp
\left( -\frac{1}{2}r^{2}-i\int^{r}z^{2\left( \ell +1\right)
}\,e^{-z^{2}}dz\right) .
\end{equation}

\item[B)] For $n_{r}=1,$ with $A_{0,\ell }^{\left( 1\right) }=1$ in (27),
the effective potential%
\begin{eqnarray}
\func{Re}V_{eff}\left( r\right) &=&\frac{\ell \left( \ell +1\right) }{r^{2}}%
+r^{2}-r^{4\left( \ell +1\right) }e^{-2r^{2}}\left( r-1\right) ^{4}  \notag
\\
&&+\frac{2\left[ \ell +1-r^{2}\right] }{r\left( r-1\right) }-2\left( \ell
+1\right) +\beta -1\text{,}
\end{eqnarray}%
\begin{equation}
\func{Im}V_{eff}\left( r\right) =4(r-1)(r^{3}-r^{2}-(\ell +2)r+\ell
+1)r^{2\ell +1}\exp (-r^{2})
\end{equation}%
which in turn leads to an $\eta $\emph{-weak-pseudo-Hermitian Hamiltonian}
of the form 
\begin{equation}
H=-\partial _{r}^{2}+\func{Re}V_{eff}\left( r\right) +i\func{Im}%
V_{eff}\left( r\right) ,
\end{equation}%
with a corresponding one-nodal (i.e., $n_{r}=1$) wave function%
\begin{eqnarray}
\psi _{1,\ell }\left( r\right) &=&N_{1,\ell }\,r^{\left( \ell +1\right)
}\left( r-1\right)  \notag \\
&&\times \exp \left( -\frac{1}{2}r^{2}-i\int^{r}z^{2\left( \ell +1\right)
}\left( z-1\right) ^{2}\,e^{-z^{2}}dz\right) .
\end{eqnarray}
\end{description}

\subsection{Perturbed radial Coulomb $\protect\eta $-weak-pseudo-Hermiticity
generator(s)}

With $\ell \geq 0$ and $\,r\in \left( 0,\infty \right) $, let the
real-valued function $g\left( r\right) =\exp \left( -2r\right) $ be
substituted in (27), then

\begin{description}
\item[A)] For $n_{r}=0,$ the effective potential%
\begin{eqnarray}
\func{Re}V_{eff}\left( r\right) &=&\frac{\ell \left( \ell +1\right) }{r^{2}}-%
\frac{2\left( \ell +1\right) }{r}-r^{4\left( \ell +1\right) }e^{-4r}+\beta +1
\notag \\
\func{Im}V_{eff}\left( r\right) &=&4\left( r-\ell -1\right) \,r^{\left(
2\ell +1\right) }\,e^{-2r}
\end{eqnarray}%
which in turn leads to an $\eta $\emph{-weak-pseudo-Hermitian Hamiltonian}
of the form 
\begin{equation}
H=-\partial _{r}^{2}+\func{Re}V_{eff}\left( r\right) +i\func{Im}%
V_{eff}\left( r\right) ,
\end{equation}%
with a corresponding node-less (i.e., $n_{r}=0$) wave function%
\begin{equation}
\psi _{0,\ell }\left( r\right) =N_{0,\ell }\,r^{\left( \ell +1\right) }\exp
\left( -r-i\int^{r}z^{2\left( \ell +1\right) }\,e^{-2z}dz\right) .
\end{equation}

\item[B)] For $n_{r}=1,$ and $A_{0,\ell }^{\left( 1\right) }=1$ in (27), the
effective potential%
\begin{eqnarray}
\func{Re}V_{eff}\left( r\right) &=&\frac{\ell \left( \ell +1\right) }{r^{2}}-%
\frac{2\left( \ell +1\right) }{r}-r^{4\left( \ell +1\right) }e^{-4r}\left(
r-1\right) ^{4}  \notag \\
&&+\frac{2\left( \ell +1\right) }{r\left( r-1\right) }-\frac{2}{r-1}-2\left(
\ell +1\right) +\beta +1\text{,}
\end{eqnarray}%
\begin{equation}
\func{Im}V_{eff}\left( r\right) =4(r-1)(r^{2}-(\ell +3)r+\ell +1)r^{2\ell
+1}e^{-2r}
\end{equation}%
which in turn leads to an $\eta $\emph{-weak-pseudo-Hermitian Hamiltonian}
of the form 
\begin{equation}
H=-\partial _{r}^{2}+\func{Re}V_{eff}\left( r\right) +i\func{Im}%
V_{eff}\left( r\right) ,
\end{equation}%
with a corresponding one-nodal (i.e., $n_{r}=1$) wave function%
\begin{eqnarray}
\psi _{1,\ell }\left( r\right) &=&N_{1,\ell }\,r^{\left( \ell +1\right)
}\left( r-1\right)  \notag \\
&&\times \exp \left( -r-i\int^{r}z^{2\left( \ell +1\right) }\left(
z-1\right) ^{2}\,e^{-2z}dz\right) .
\end{eqnarray}
\end{description}

\section{Concluding Remarks}

In this work, we have presented a class of spherically symmetric
non-Hermitian Hamiltonians and their $\eta $-\emph{weak-pseudo-Hermiticity
generators. We have used }an operators-based procedure to come out with $%
\eta $-\emph{weak-pseudo-Hermitian }non-Hermitian weakly perturbed 1D and
radial harmonic oscillators, perturbed radial Coulomb, and the radial Morse
models. We have presented not only nodeless 1D and radial wave functions but
also some of the multi-nodal ones.

In the light of this experience, we have witnessed that the form of an $\eta 
$-\emph{weak-pseudo-Hermitian Hamiltonian} changes for different values of $%
\ell $ and $n_{r}$. However, the reader should be reminded that our current
proposal does not target exact solvability of $\eta $-\emph{%
weak-pseudo-Hermitian Hamiltonians} but rather produces non-Hermitian $\eta $%
-\emph{weak-pseudo-Hermitian Hamiltonians} with real spectra. Consequently,
the Hermiticity requirement to ensure the reality of the spectrum of a
Hamiltonian is shown to be not only fragile but also mathematically
unnecessarily strong. Nevertheless, more physically oriented (although in
1D-case) applications of the above procedure could be found in our recent
work in [16]. Therein, we have explored just a possibility of exact
solvability through a Scarf II and a periodic-type $\eta $-\emph{%
weak-pseudo-Hermitian Hamiltonian }models. Yet, another option for the
intertwining operator $\eta $ as a first-order intertwiner is explored in
[17] with few informative examples presented.

\end{document}